\documentclass[submission,copyright,creativecommons]{eptcs}
\providecommand{\event}{ACL2 Workshop 2020} % Name of the event you are submitting to
\usepackage{amsmath,amssymb,amsfonts,amsthm}
\usepackage{fancyvrb}
\usepackage{alltt}
\usepackage[shortcuts]{extdash}
\usepackage{color}
\usepackage{framed}
\usepackage{xspace}
\usepackage{listings}
\usepackage{tabto}
\newcommand{\loopc}{\texttt{loop}\xspace}
\newcommand{\loopd}{\texttt{loop\$}\xspace}
\newcommand{\Loopd}{\texttt{Loop\$}\xspace}
\newcommand{\loopdas}{\texttt{loop\$-as}\xspace}
\newcommand{\Loopdas}{\texttt{Loop\$-as}\xspace}
\newcommand{\applyd}{\texttt{apply\$}\xspace}
\newcommand{\Applyd}{\texttt{Apply\$}\xspace}
\newcommand{\collectd}{\texttt{collect\$}\xspace}
\newcommand{\alwaysd}{\texttt{always\$}\xspace}
\newcommand{\thereisd}{\texttt{thereis\$}\xspace}
\newcommand{\appendd}{\texttt{append\$}\xspace}
\newcommand{\sumd}{\texttt{sum\$}\xspace}
\newcommand{\Sumd}{\texttt{Sum\$}\xspace}
\newcommand{\defund}{\texttt{defun\$}\xspace}
\newcommand{\whend}{\texttt{when\$}\xspace}
\newcommand{\untild}{\texttt{until\$}\xspace}
\newcommand{\lambdad}{\texttt{lambda\$}\xspace}

\newcommand{\collectdp}{\texttt{collect\$+}\xspace}
\newcommand{\alwaysdp}{\texttt{always\$+}\xspace}
\newcommand{\thereisdp}{\texttt{thereis\$+}\xspace}
\newcommand{\appenddp}{\texttt{append\$+}\xspace}
\newcommand{\sumdp}{\texttt{sum\$+}\xspace}

\newcommand{\whendp}{\texttt{when\$+}\xspace}
\newcommand{\untildp}{\texttt{until\$+}\xspace}
\newcommand{\nil}{\texttt{nil}\xspace}

\newcommand{\kALWAYS}{\texttt{ALWAYS}\xspace}
\newcommand{\kTHEREIS}{\texttt{THEREIS}\xspace}
\newcommand{\kAPPEND}{\texttt{APPEND}\xspace}
\newcommand{\kAS}{\texttt{AS}\xspace}
\newcommand{\kCOLLECT}{\texttt{COLLECT}\xspace}
\newcommand{\kFOR}{\texttt{FOR}\xspace}
\newcommand{\kSUM}{\texttt{SUM}\xspace}
\newcommand{\kUNTIL}{\texttt{UNTIL}\xspace}
\newcommand{\kWHEN}{\texttt{WHEN}\xspace}

\newcommand{\hrefdoc}[2]{\href{http://www.cs.utexas.edu/users/moore/acl2/manuals/current/manual/index.html?topic=#1}{#2}}

\newcommand{\hrefdocu}[2]{\hrefdoc{#1}{{\underline{#2}}}}
\newcommand{\hrefdocutt}[2]{\hrefdocu{#1}{\texttt{#2}}}

\definecolor{darkcyan}{cmyk}{1.0,0.2,0.2,0.2}

\title{Iteration in ACL2}
\author{Matt Kaufmann and J Strother Moore
\institute{Department of Computer Science\\
The University of Texas at Austin, Austin, TX, USA}
\email{\{kaufmann,moore\}@cs.utexas.edu}
}
\def\titlerunning{Loop}
\def\authorrunning{M. Kaufmann \& J S. Moore}
\begin{document}
\maketitle

\begin{abstract}

Iterative algorithms are traditionally expressed in ACL2 using
recursion.  On the other hand, Common Lisp provides a construct,
\loopc, which --- like most programming languages --- provides direct
support for iteration.  We describe an ACL2 analogue \loopd of \loopc
that supports efficient ACL2 programming and reasoning with iteration.

\end{abstract}

\section{Introduction}
\label{sec:intro}

Recursion is a natural way to define notions in an equational logic.
But for programming, iteration is often more natural and concise than
recursion.

This paper introduces an ACL2 iteration construct, \loopd.  We give an
overview of its usage in programming and reasoning, and we touch on
some key aspects of its implementation.  Those who want more complete
user-level documentation are welcome to see :DOC
\hrefdoc{ACL2\_\_\_\_LOOP\_42}{loop\$}~\footnote{In the online version
  of this paper, we follow the usual convention of adding links to
  documentation on the web~\cite{acl2-plus-books-manual}.  Sometimes we say ``see
  :DOC {\it topic}'', but other times we simply underline a
  \hrefdocu{COMMON-LISP\_\_\_\_DOCUMENTATION}{documentation} topic.},
and those who want more implementation details may visit Lisp comments
in the ACL2 sources, in particular the ``Essay on Loop\$'' and the
``Essay on Evaluation of Apply\$ and Loop\$ Calls During Proofs''.

The rest of this introduction illustrates the basics of \loopd and
then mentions some related prior work.  Next, Section~\ref{sec:syntax}
outlines the syntax of \loopd expressions.  Section~\ref{sec:warrants}
provides necessary background on \applyd and especially {\em
  warrants}.  Then Section~\ref{sec:logic} gives semantics to \loopd
expressions by translating them into the logic, which provides
background for the discussion of reasoning about \loopd in
Section~\ref{sec:reasoning}.  Section~\ref{sec:evaluation} discusses
evaluation of calls of \loopd, including considerations involving
guards.  We conclude by considering possible future enhancements to
\loopd.

The \hrefdoc{ACL2\_\_\_\_COMMUNITY-BOOKS}{community book}
\texttt{projects/apply/loop-tests.lisp} constitutes supporting
materials for this paper: it contains most ACL2 events and expressions
discussed below.

\subsection{\Loopd basics}

Common Lisp supports iteration with the construct \loopc, and ACL2
provides the analogous (though less general) construct
\loopd.\footnote{The dollar sign is commonly used as a suffix in ACL2
  to distinguish from a similar Common Lisp utility.  Note that the
  symbol {\loopc} in the \texttt{"COMMON-LISP"} package is one of the
  978 external symbols in that package that are imported into the
  \texttt{"COMMON-LISP"} package, so we cannot distinguish
  \texttt{acl2::loop} from \texttt{common-lisp::loop}; these are the
  same symbol.}  These two forms evaluate to 30, in Common Lisp and in
ACL2, respectively.

\begin{verbatim}
(loop for x in '(1 2 3 4) sum (* x x))
(loop$ for x in '(1 2 3 4) sum (* x x))
\end{verbatim}

A call of \loopd is given semantics by translating it to a logical
{\em translated \hrefdocu{ACL2\_\_\_\_TERM}{term}} that is a call of a
% , extending the usual notion of translation to logic (see
% :DOC \hrefdocu{ACL2\_\_\_\_TERM}{term}); in particular, macro calls are
% expanded, e.g., \texttt{(* x x)} is expanded to the function call,
% \texttt\texttt{(binary-* x x)}.  \Loopd calls are translated into
% calls of {\em \loopd scions}, which are something like higher-order
{\em \loopd scion}, in the spirit of a higher-order function call.
For example the translation of the \loopd form above is the following
call of the \loopd scion, \sumd, whose first argument is a quoted {\em
  \hrefdocu{COMMON-LISP\_\_\_\_LAMBDA}{lambda} object}.

\begin{verbatim}
(SUM$ '(LAMBDA (X) (BINARY-* X X))
      '(1 2 3 4))
\end{verbatim}

\noindent \Sumd has the following definition, ignoring declarations
and calls of \hrefdocutt{ACL2\_\_\_\_MBE}{mbe} and
\hrefdocutt{ACL2\_\_\_\_FIX}{fix}.

\begin{verbatim}
(defun sum$ (fn lst)
  (if (endp lst)
      0
    (+ (apply$ fn (list (car lst)))
       (sum$ fn (cdr lst)))))
\end{verbatim}

\noindent Notice the use of \hrefdocu{ACL2\_\_\_\_APPLY\_42}{\applyd},
to apply a given function symbol or
\hrefdocu{COMMON-LISP\_\_\_\_LAMBDA}{lambda} object to a list of
arguments.  Support for \loopd thus depends on the sort of
``higher-order'' capability provided by \applyd.  (We return to
\applyd in Section~\ref{sec:warrants}.)

In the following example we use a different \loopd scion, \collectd,
to collect into a list instead of summing results.  Also, we filter by
considering only even numbers that are multiples of 5 (thus: multiples
of 10), stopping when we hit a number greater than 30.

\begin{verbatim}
ACL2 !>(loop$ for i from 0 to 1000000 by 5
              until (> i 30)
              when (evenp i) collect (* i i))
(0 100 400 900)
ACL2 !>
\end{verbatim}

\noindent This time we get essentially the following translation of
the \loopd form.  Reading it bottom-up (inside-out), we see that
\texttt{from-to-by} builds a list of integers \texttt{(0 5 10 15
  $\ldots$ 1000000)}, which \untild reduces to \texttt{(0 5 10 15 20
  25 30)} by stopping when \texttt{30} is exceeded; then \whend
restricts that list to the even integers to produce the list
\texttt{(0 10 20 30)}, and finally \collectd assembles the squares of
the members of that list to produce the result, \texttt{(0 100 400
  900)}.  We believe that this compositional, though computationally inefficient, formal
semantics is useful for reasoning about applications of \loopd since
lemmas can be proved to simplify the respective steps.

\begin{verbatim}
(COLLECT$ '(LAMBDA (I) (BINARY-* I I))
          (WHEN$ '(LAMBDA (I) (EVENP I))
                 (UNTIL$ '(LAMBDA (I) (< '30 I))
                         (FROM-TO-BY '0 '1000000 '5))))
\end{verbatim}

\hrefdocu{ACL2\_\_\_\_LOOP\_42}{Loop\$} scions (including \collectd,
\whend, and \untild) repeatedly call
\hrefdocu{ACL2\_\_\_\_APPLY\_42}{apply\$} on a given lambda object (or
function symbol).  In Section~\ref{sec:warrants} we explore
consequences of these uses of \applyd.

If we define a \hrefdocu{ACL2\_\_\_\_GUARD}{guard}-verified function
using a corresponding call of \loopd, then evaluation is more
efficient.  Consider the following variation of the example just
above, which adds an \texttt{of-type} clause to aid in guard
verification (more on that is in Section~\ref{sec:evaluation}).  The
\hrefdocutt{ACL2\_\_\_\_INCLUDE-BOOK}{include-book} event shown is
important for almost all work that involves \applyd or \loopd.

\begin{verbatim}
ACL2 !>(include-book "projects/apply/top" :dir :system)
[[.. output omitted ..]]
ACL2 !>(defun f1 ()
         (declare (xargs :guard t))
         (loop$ for i of-type integer from 0 to 1000000 by 5
                until (> i 30)
                when (evenp i) collect (* i i)))
[[.. output omitted ..]]
ACL2 !>(f1)
(0 100 400 900)
ACL2 !>
\end{verbatim}

\noindent After \texttt{f1} is admitted with guards verified, its
Common Lisp definition is used for evaluation (as usual).  The Common
Lisp macroexpansion of the \loopd form above is essentially as
follows, which presumably allows the Common Lisp compiler to generate
efficient code, in particular avoiding the indirection of \applyd,
the repeated passes implied by our compositional semantics, and
(presumably) use of the heap to form the intermediate lists noted
above.

\begin{verbatim}
(LOOP FOR I OF-TYPE INTEGER FROM 0 TO 1000000 BY 5
      UNTIL (> I 30)
      WHEN (EVENP I) COLLECT (* I I))
\end{verbatim}

\subsection{Before ACL2}

The addition of an iteration construct within the first-order logic of
the Boyer-Moore provers has been a longstanding goal.  It was first
explored by Moore in 1975~\cite{iteration}.  The Nqthm
prover~\cite{aclh2} of Boyer and Moore provided an iterative
construct, called \texttt{FOR}, which provided simple loops such as
% Do not use \kFOR for \texttt{FOR} in this subsection, since that's a
% reference to an NQTHM construct, not to the loop$ keyword.

\begin{verbatim}
(FOR 'X (APPEND U V)
     '(MEMBER X B)
     'COLLECT '(TIMES X C)
     (LIST (CONS 'B B) (CONS 'C C))).
\end{verbatim}

\noindent The expression above can be written in Common Lisp as

\begin{verbatim}
(loop for x in (append u v) when (member x b) collect (* x c)).
\end{verbatim}

Nqthm's \texttt{FOR} was defined using a universal interpreter for
Nqthm called \texttt{V\&C\$} and many familiar theorems about
\texttt{FOR} were proved by Nqthm~\cite{bounded-quant}.  However,
\texttt{V\&C\$} was a non-constructive
function~\cite{Kunen:1998:NCM:594117.594175} and, in addition, was not
compatible with \hrefdocu{ACL2\_\_\_\_LOCAL}{local} events; Nqthm,
unlike ACL2, did not support \texttt{local}.  Thus \texttt{V\&C\$} was
abandoned for ACL2.
% [From Matt] I found ``non-constructive'' a bit misleading, as it
% seemed to suggest that evaluation wasn't supported, yet I was able
% to evaluate (foo nil '(3) '(3) 7) inside (r-loop) to get the
% expected result of '(21).  However, I didn't change anything,
% because ``non-constructive'' is accurate and I didn't really want to
% expand this section.

\section{Syntax}\label{sec:syntax}

The syntax of \loopd is a restriction of the syntax of Common Lisp's
\loopc macro:

\begin{alltt}
(LOOP\$ FOR \(v\sb{1}\) OF-TYPE \(s\sb{1}\) \(tgt\sb{1}\)      {\em{; where}} OF-TYPE \(s\sb{1}\) {\em{is optional}}
{\em{; The following}} AS {\em{clauses are optional.}}
       AS  \(v\sb{2}\) OF-TYPE \(s\sb{2}\) \(tgt\sb{2}\)      {\em{; where}} OF-TYPE \(s\sb{2}\) {\em{is optional}}
\ldots
       AS  \(v\sb{n}\) OF-TYPE \(s\sb{n}\) \(tgt\sb{n}\)      {\em{; where}} OF-TYPE \(s\sb{n}\) {\em{is optional}}
{\em{; The following}} UNTIL {\em{and}} WHEN {\em{clauses are optional.}}
       UNTIL :GUARD \(g\sb{u}\) \(u\)     \(\,\)    {\em{; where}} :GUARD \(g\sb{u}\) {\em{is optional}}
       WHEN  :GUARD \(g\sb{w}\) \(w\)     \(\,\,\:\)   {\em{; where}} :GUARD \(g\sb{w}\) {\em{is optional}}
       op    :GUARD \(g\sb{b}\) \(b\)     \(\:\:\:\,\,\)   {\em{; where}} :GUARD \(g\sb{b}\) {\em{is optional}}
       )
\end{alltt}

\noindent where the $v_i$ are distinct variables, the {\em iteration
  variables}; each \texttt{OF-TYPE $s_i$} is optional where $s_i$ is a
\hrefdocu{ACL2\_\_\_\_TYPE-SPEC}{type-spec}; each $tgt_i$ is a {\em
  target clause} (see below); each of $g_u$, $u$, $g_w$, $w$, $g_b$,
and $b$ is a term; and \texttt{op} is one of the {\em loop operators}
\kALWAYS, \kTHEREIS, \kAPPEND, \kCOLLECT, or \kSUM.  We call $b$ the {\em loop
  body}.  See :DOC \hrefdoc{ACL2\_\_\_\_LOOP\_42}{loop\$} for
restrictions, e.g., on terms being in
\hrefdocu{ACL2\_\_\_\_LOGIC}{\texttt{:logic}} mode and
\hrefdocu{ACL2\_\_\_\_TAME}{\textit{tame}} --- where user-defined
function symbols all have warrants (see Section~\ref{sec:warrants}) --- and
on not allowing both a \kWHEN clause and an \kALWAYS or \kTHEREIS operator.

A {\em target clause} has one of the following forms, where $lst$ is a
term denoting a list, $lo$ and $hi$ are terms denoting integers, and
the {\em step expression} $s$ (if supplied) is a term denoting a
positive integer.

\begin{itemize}
\item \texttt{IN} $lst$
\item \texttt{ON} $lst$
\item \texttt{FROM} $lo$ \texttt{TO} $hi$
\item \texttt{FROM} $lo$ \texttt{TO} $hi$ \texttt{BY} $s$
\end{itemize}

\noindent The corresponding {\em target list} is $lst$ in the first case,
the list of non-empty tails of $lst$ in the second, 
the list of integers {\tt{($lo$ $lo+1$ $\ldots$ $hi$)}} in the
  third case, and the list of integers {\tt{($lo$ $lo+s$ $\ldots$)}}
  terminating just before exceeding $hi$ in the fourth case.

\section{\Applyd and Warrants}\label{sec:warrants}

All of the \loopd scions are defined using calls of \applyd.  Thus,
the value of a \loopd expression depends on values of \applyd calls.
If $F$ is a function symbol of arity $n$, then it is clearly desirable
for both rewriting and evaluation to replace a call
\texttt{(\applyd\ '$F$ (list $t_1 \ldots t_n$))} by \texttt{($F$ $t_1
  \ldots t_n$)}.  But for user-defined $F$, the equality of these two
terms depends on the {\em warrant hypothesis} for $F$: the assertion
\texttt{(apply\$-warrant-$F$)}, where \texttt{apply\$-warrant-$F$} is
the \hrefdocu{ACL2\_\_\_\_WARRANT}{warrant} of $F$.  The interested reader may
find further background on \applyd, including logical issues about the
need for warrants, in our paper on \applyd~\cite{apply-jar} and in
discussions in :DOC \hrefdoc{ACL2\_\_\_\_APPLY\_42}{\texttt{apply\$}}, :DOC
\hrefdoc{ACL2\_\_\_\_WARRANT}{warrant}, and (regarding the so-called
``local problem'') :DOC
\hrefdoc{ACL2\_\_\_\_INTRODUCTION-TO-APPLY\_42}{introduction-to-apply\$}.

The example below illustrates warrants and warrant hypotheses.  We
begin as follows where as noted above, the
\hrefdocutt{ACL2\_\_\_\_INCLUDE-BOOK}{include-book} event is important
for almost all work that involves \applyd or \loopd.

\begin{verbatim}
(include-book "projects/apply/top" :dir :system)

(defun$ square (n)
  (declare (xargs :guard (integerp n)))
  (* n n))
\end{verbatim}

\noindent A call of \hrefdocu{ACL2\_\_\_\_DEFUN\_42}{\defund}
generates a corresponding \hrefdocutt{COMMON-LISP\_\_\_\_DEFUN}{defun}
event and introduces the warrant for \texttt{square}, which is the
function symbol \texttt{apply\$-warrant-square}, with the event
\texttt{(defwarrant square)}.  The following rewrite rule, named
\texttt{apply\$-square}, is the key property of the warrant
hypothesis, which is \hrefdocu{ACL2\_\_\_\_FORCE}{force}d here so that
a proof may progress to the forcing round as described below.

\begin{verbatim}
(implies (force (apply$-warrant-square))
         (equal (apply$ 'SQUARE args)
                (square (car args))))
\end{verbatim}

\noindent We may abbreviate \texttt{(apply\$-warrant-square)} with the
macro call, \texttt{(warrant square)}.  More generally, for a function
symbol $F$, \texttt{(warrant $F$)} abbreviates
\texttt{(apply\$-warrant-$F$)}.  We see that the rewrite rule above
replaces a call of \applyd on \texttt{'SQUARE} by the corresponding
call of \texttt{square}.

In summary: the warrant hypothesis for $F$
justifies replacing \texttt{(\applyd\ '$F$ (list $t_1 \ldots t_n$))} by
\texttt{($F$ $t_1 \ldots t_n$)}.

The following sequence of events, extending the \texttt{include-book}
and \defund forms displayed above, illustrates the role of warrant
hypotheses.

\begin{Verbatim}[commandchars=\\\{\}]
{\em{; Return the list of squares of integers between the given bounds.}}
(defun f2 (lower upper)
  (declare (xargs :guard (and (integerp lower) (integerp upper))))
  (loop$ for i of-type integer from lower to upper
         collect (square i)))

(assert-event (equal (f2 3 5) '(9 16 25))) {\em{; example evaluation}}

(thm (implies (warrant square)
              (equal (f2 3 5) '(9 16 25))))

(thm (implies (and (natp k1) (natp k2) (natp k3) (<= k1 k2) (<= k2 k3)
                   (warrant square))
              (member (* k2 k2) (f2 k1 k3))))
\end{Verbatim}

If the warrant hypothesis is omitted in the two calls of \texttt{thm}
above, the proofs will progress to forcing rounds before failing.  The
forced goals make clear the need for a warrant hypothesis.  For
example, if the warrant hypothesis is removed before submitting the
first \texttt{thm} form above, then the forcing round has the goal
\texttt{(APPLY\$-WARRANT-SQUARE)}.\footnote{Missing warrants from
otherwise provable theorems do not {\it{necessarily}} lead to forcing
because the proof may fail before then.}

Note that proofs succeed automatically for both \texttt{thm} forms
above.  They succeed without the initial \texttt{include-book} event
and without any warrant hypotheses if we replace \texttt{(square x)}
by \texttt{(* x x)} in the definition of \texttt{f2}; only
user-defined functions have warrants (or require warrant hypotheses),
not primitives like \texttt{*} (more precisely, \texttt{binary-*};
\texttt{*} is a macro).

\section{Semantics: Translation to Logic}\label{sec:logic}

The introduction presented the example

\begin{verbatim}
(loop$ for x in '(1 2 3 4) sum (* x x))
\end{verbatim}

\noindent and stated that it ``essentially'' translates to the
following term.

\begin{verbatim}
(SUM$ '(LAMBDA (X) (BINARY-* X X))
      '(1 2 3 4))
\end{verbatim}

\noindent
In this section we explain precisely how the simplest, {\em plain}
\loopd expressions such as this one are translated into \loopd scion
calls.  We then discuss the translation of more complex {\em fancy}
\loopd expressions.  We conclude with a note about relaxation of some
restrictions for \loopd expressions in theorems.

\subsection{Plain loops and \lambdad}\label{sec:plain-loops}

A {\em plain loop} is a \loopd expression that has a single iteration
variable --- that is, a \kFOR clause but no \kAS clause --- which is
the only variable that occurs free in the \kUNTIL test, \kWHEN test, or loop body.  ACL2 gives
semantics to a plain loop by generating a call of the \loopd scion
corresponding to the loop operator, where the first argument is a {\em
  \hrefdocu{ACL2\_\_\_\_LAMBDA\_42}{\lambdad} expression} whose single
formal is the iteration variable and whose body is the loop body.  A
\lambdad expression, in turn, represents a corresponding quoted
\hrefdocu{COMMON-LISP\_\_\_\_LAMBDA}{lambda} object whose body is a
translated term.  Finally, additional calls of
\hrefdocutt{ACL2\_\_\_\_RETURN-LAST}{return-last} are inserted
(see below).

Let's see how this works on the \loopd expression in the following definition.

\begin{verbatim}
(defun sum-squares (lst)
  (loop$ for x in lst sum (* x x)))
\end{verbatim}

\noindent This is a simple loop: it binds only the variable
\texttt{x}, which is the only variable occurring free in the
(translation of) the loop body.  ACL2 first transforms this \loopd
expression to

% Use (trace$ translate11) to see this.

\begin{verbatim}
(SUM$ (LAMBDA$ (X) (* X X))
      LST).
\end{verbatim}

\noindent We see that the loop operator \kSUM generates a call of the
corresponding \loopd scion, \sumd.  The first argument of that call is
the \lambdad expression whose single formal is the iteration variable,
\texttt{x}, and whose body is exactly the loop body.  This \lambdad
expression, in turn, is transformed to a quoted lambda object by
replacing the loop body \texttt{(* x x)} with its translation
\texttt{(binary-* x x)}, and by declaring that the formals --- in this
case, the single formal, \texttt{x} --- may be ignored.  (Such a
declaration permits loop bodies that do not mention all of the iteration
variables.)

\begin{verbatim}
(SUM$ '(LAMBDA (X)
               (DECLARE (IGNORABLE X))
               (BINARY-* X X))
      LST)
\end{verbatim}

\noindent But we're not done yet!  The body of the \lambdad expression
is actually replaced by the translation of
\texttt{(\hrefdocu{ACL2\_\_\_\_PROG2\_42}{prog2\$} '(LAMBDA\$ (X) (* X
  X)) (* x x))} --- the system uses that quoted \lambdad expression
for efficient execution of compiled lambda objects\footnote{See ACL2
  source file \texttt{apply-raw.lisp} for details, in particular, the
  ``Essay on the CL-Cache Implementation Details''.} --- and the
\loopd expression similarly is wrapped with a call of \texttt{prog2\$}
that preserves the original \loopd expression in the first argument,
as discussed below in Subsection~\ref{sec:guards}.  Here is the
translation of the body of \texttt{sum-squares}.

\begin{verbatim}
(RETURN-LAST 'PROGN
             '(LOOP$ FOR X IN LST SUM (* X X))
             (SUM$ '(LAMBDA (X)
                            (DECLARE (IGNORABLE X))
                            (RETURN-LAST 'PROGN
                                         '(LAMBDA$ (X) (* X X))
                                         (BINARY-* X X)))
                   LST))
\end{verbatim}

Fortunately, one does not generally see such complexity during a
proof.  One reason is that the \texttt{return-last} calls are removed;
see :DOC \hrefdoc{ACL2\_\_\_\_GUARD-HOLDERS}{guard-holders}.  Another
reason is that during a proof one sees {\em untranslated terms} (see
:DOC \hrefdoc{ACL2\_\_\_\_TERM}{term}), and the untranslation process
typically restores the original \lambdad expression, so that in this
case one sees \texttt{(sum\$ (lambda\$ (x) (* x x)) lst)}.

We remark that a \lambdad expression is only legal when, roughly
speaking, it is ultimately used only as the first argument of \applyd
calls.\footnote{To be precise, it must be in a position of {\em ilk}
  \texttt{:FN}~\cite{apply-jar}.}

\subsection{Fancy loops}\label{sec:fancy-loops}

We explain translation of fancy loops using the following example.
This definition contains a fancy loop both because the variables
\texttt{m} and \texttt{n} are not bound by the \loopd expression and
because of the \kAS clause.

\begin{verbatim}
(defun g (m n lst1 lst2)
  (loop$ for x1 in lst1 as x2 in lst2 sum (* m n x1 x2)))
\end{verbatim}

\noindent For our discussion of fancy loops we will focus on basic semantics,
ignoring \texttt{return-last} calls and \texttt{ignorable}
declarations discussed for plain loops above.

Before we show the (essential) translation of the fancy loop above, we
introduce auxiliary functions, \loopdas and \sumdp.  \Loopdas is given
a list \texttt{L} of $n$ lists and returns the sequence of $n$-tuples containing
corresponding elements from each list until the shortest list is exhausted.
For example:

\begin{verbatim}
ACL2 !>(loop$-as '((1 2 3 4 A B C) (5 6 7 8)))
((1 5) (2 6) (3 7) (4 8))
ACL2 !>
\end{verbatim}

We next consider the {\em fancy \loopd scion} \sumdp, which is called
in the translation of the fancy loop above.  Its first parameter, \texttt{fn}, is
expected to be a function symbol or lambda object with two arguments.  The first
argument of \texttt{fn} is expected to be a list of {\em globals}:
values of the variables in the loop body that are not iteration
variables.  The second argument of \texttt{fn} is expected to be a
list of {\em locals}: the result of a \loopdas call, which (again) is
a list whose $n^{th}$ member lists the $n^{th}$ iteration values of the
iteration variables.  Thus \sumdp is defined much like \sumd
(here we ignore declarations as well as \texttt{mbe} and \texttt{fix}
wrappers), but where the globals are passed when applying \texttt{fn}
and the locals are wrapped into a list.

\begin{verbatim}
(defun sum$+ (fn globals lst)
  (if (endp lst)
      0
    (+ (apply$ fn (list globals (car lst)))
       (sum$+ fn globals (cdr lst)))))
\end{verbatim}

The fancy loop in the definition of \texttt{g}, above, has the
following semantics (translation).  We see that the first argument
is indeed a function of globals and locals, where the globals is the
list containing the values of formals \texttt{m} and \texttt{n} of the
function \texttt{g} defined above and the locals come from the
\loopdas call.

\begin{Verbatim}[commandchars=\\\{\}]
(SUM$+ (LAMBDA$ (LOOP$-GVARS LOOP$-IVARS)
                (DECLARE (XARGS :GUARD ...))
                (LET ((M (CAR LOOP$-GVARS))
                      (N (CAR (CDR LOOP$-GVARS)))
                      (X1 (CAR LOOP$-IVARS))
                      (X2 (CAR (CDR LOOP$-IVARS))))
                     (* M N X1 X2)))
       (LIST M N)
       (LOOP$-AS (LIST LST1 LST2)))
\end{Verbatim}

We have seen that the \loopd operator, \kSUM, produces a call of a
\loopd scion: \sumd for a plain loop and \sumdp for a fancy loop.
More generally, a \loopd scion arises from any use of \loopd, and the scions introduced
are all plain or all fancy depending on the particulars of the \loopd.
The plain
ones are \sumd, \collectd, \alwaysd, \thereisd, \appendd, \untild, and \whend,
respectively, and their fancy counterparts are \sumdp, \collectdp,
\alwaysdp, \thereisdp, \appenddp, \untildp, and \whendp.  The plain \loopd scions
may be summarized as follows, where the elements of the true-list
\texttt{lst} are $e_1, \ldots, e_n$.

\begin{itemize}

\item \texttt{(\sumd fn lst)}: sums all \texttt{(\applyd fn (list
  $e_i$))}

\item \texttt{(\alwaysd fn lst)}: tests that all \texttt{(\applyd fn
  (list $e_i$))} are non-\nil

\item \texttt{(\thereisd fn lst)}: returns the first non-\nil value of
  \texttt{(\applyd fn (list $e_i$))}, $i=1,2,\ldots,n$
% [From Matt to J] We could consider omitting ``, $i=1,2,\ldots,n$''
% just above, since we don't have any such wording in the other items.
% But I left it there in case you wanted to emphasize the order.

\item \texttt{(\collectd fn lst)}: conses together all
  \texttt{(\applyd fn (list $e_i$))}

\item \texttt{(\appendd fn lst)}: appends together all
  \texttt{(\applyd fn (list $e_i$)})

\item \texttt{(\untild fn lst)}: lists all $e_i$, in order, until the 
  first $i$ such that \texttt{(\applyd fn (list $e_i$))} is non-\nil

\item \texttt{(\whend fn lst)}: lists those $e_i$, in order, such that
  \texttt{(\applyd fn (list $e_i$))} is non-\nil

\end{itemize}

The fancy \loopd scions are analogous.  We have already seen the
definition of \sumd, and we have seen a similar definition of \sumdp
that accommodates the globals.  Here is an analogous pair of definitions
for \whend and \whendp.

\begin{verbatim}
(defun when$ (fn lst)
  (if (endp lst)
      nil
    (if (apply$ fn (list (car lst)))
        (cons (car lst) (when$ fn (cdr lst)))
      (when$ fn (cdr lst)))))

(defun when$+ (fn globals lst)
  (if (endp lst)
      nil
    (if (apply$ fn (list globals (car lst)))
        (cons (car lst) (when$+ fn globals (cdr lst)))
      (when$+ fn globals (cdr lst)))))
\end{verbatim}

%% [J agrees that it's not necessary to say more as in the possible
%% extra subsection mentioned below.]  We could show translations in
%% FROM-TO-BY and ON cases. Anything else?  Maybe :GUARD with forward
%% pointer to guards section.  But actually, I think that such a
%% subsection is not necessary, so I've tentatively commented it out
%% below.  Thoughts?

% \subsection{Some other forms of \loopd}

% For more complete information on the legal forms of \loopd
% expressions, see :DOC \hrefdoc{ACL2\_\_\_\_LOOP\_42}{loop\$}.

\subsection{\Loopd restrictions are eased in theorems}

% This subsection can be deleted if necessary.

Conventional syntactic requirements are enforced for the loop body of
a \loopd expression that occurs either at the top level or in the body
of a definition.  In particular, the loop body should return a single,
non-\hrefdocu{ACL2\_\_\_\_STOBJ}{stobj} value.  As usual, such requirements are relaxed when the
\loopd expression occurs in a theorem.  For example, the following
form is legal, and in fact is admitted by ACL2, even though the \loopd
expression is illegal both at the top level and in any function body.

\begin{verbatim}
(thm (equal (loop$ for x in '(A B C) collect (mv x x))
            '((A A) (B B) (C C))))
\end{verbatim}

\section{Reasoning with \Loopd}\label{sec:reasoning}

We recall the definition of \texttt{sum-squares} above and state a
simple theorem about that function.

\begin{verbatim}
(defun sum-squares (lst)
  (loop$ for x in lst sum (* x x)))

(thm (equal (sum-squares (reverse x))
            (sum-squares x)))
\end{verbatim}

\noindent The proof fails, but the experienced ACL2 user quickly
completes the proof by noticing terms of the form \texttt{(\sumd FN
  (revappend X Y))} in the checkpoints and then proving the following
lemma.

\begin{verbatim}
(defthm sum$-revappend
  (equal (sum$ fn (revappend x y))
         (+ (sum$ fn x) (sum$ fn y))))
\end{verbatim}

Of course, one could instead define \texttt{sum-squares} recursively
and go through a similar process to find and prove a suitable lemma
about \texttt{(sum-squares (revappend x y))}, and then prove the
\texttt{thm}.  But by using \loopd, one needn't prove such a lemma
more than once.  For example, suppose we define:

\begin{verbatim}
(defun sum-cubes (lst)
  (loop$ for x in lst sum (* x x x)))
\end{verbatim}

\noindent Then the following theorem is proved automatically by
applying the lemma above, \texttt{sum\$-revappend}.

\begin{verbatim}
(thm (equal (sum-cubes (reverse x))
            (sum-cubes x)))
\end{verbatim}

\noindent If instead \texttt{sum-cubes} were defined by recursion,
then one would need first to prove a new lemma about
\texttt{(sum-cubes (revappend x y))}.

We have seen that reasoning about \loopd expressions generally reduces
to reasoning about \loopd scions.  The
\hrefdoc{ACL2\_\_\_\_COMMUNITY-BOOKS}{community book}
\texttt{projects/apply/loop}, included in the
book \texttt{projects/apply/top} that is typically included
when reasoning about \applyd or \loopd, has helpful lemmas to support
automation of reasoning about \loopd.  Perhaps over time, the lemma
\texttt{sum\$-revappend} (above) and other useful lemmas will be
included in lemma libraries for reasoning about \loopd.  Also see
community book \texttt{projects/apply/report.lisp} for examples of
reasoning with (close analogues of) \loopd scions.

Let's look at one more proof example.

\begin{verbatim}
(thm (equal (sum-squares '(1 2 3 4)) 30))
\end{verbatim}

\noindent This proves automatically using these rules.

\begin{verbatim}
Rules: ((:EXECUTABLE-COUNTERPART EQUAL)
        (:EXECUTABLE-COUNTERPART SUM-SQUARES))
\end{verbatim}

\noindent Thus, evaluation involving \loopd expressions (in this case,
in the body of the definition of \texttt{sum-squares}) can take place
during proofs.  This simple example does not require warrants, but
those are handled with evaluation as well; see
Section~\ref{sec:eval-proofs}.

The \hrefdoc{ACL2\_\_\_\_COMMUNITY-BOOKS}{community book}
\texttt{projects/apply/mempos.lisp} may be helpful for reasoning about
\loopd expressions that have more than one iteration variable or an
\kUNTIL clause.

\section{Evaluation}\label{sec:evaluation}

We have seen that \loopd calls are translated into logic by generating
calls of \loopd scions.  Thus, evaluation of \loopd calls can take
place by evaluating those translations.  Indeed, that is what happens
when \loopd expressions occur at the top level, as we discuss in the
first subsection below.  However, when \loopd expressions are in
guard-verified code, their evaluation generally reduces to efficient
evaluation of corresponding Common Lisp \loopc expressions, as we
discuss in the second subsection below.  We next discuss a key
exception: during proofs, \loopd scions are always used for
evaluation, with a twist: there is accounting for warrants.  Finally
we discuss performance.

\subsection{Evaluation in the top-level loop}\label{eval-using-scions}

We first consider evaluation of \loopd calls taking place directly in
the top-level loop (rather than within a function body).  Every
top-level form is evaluated by first translating into logic.  In
particular, evaluation of a \loopd call in the top-level loop takes
place by translating into logic, that is, into calls of \loopd scions
as described in Section~\ref{sec:logic}.  The following example shows
this behavior in action.

\begin{verbatim}
ACL2 !>(trace$ collect$)
 ((COLLECT$))
ACL2 !>(loop$ for i from 1 to 5 collect (* i i))
1> (ACL2_*1*_ACL2::COLLECT$
        (LAMBDA (I)
                (DECLARE (IGNORABLE I))
                (RETURN-LAST 'PROGN
                             '(LAMBDA$ (I) (* I I))
                             (BINARY-* I I)))
        (1 2 3 4 5))
  2> (COLLECT$ (LAMBDA (I)
                       (DECLARE (IGNORABLE I))
                       (RETURN-LAST 'PROGN
                                    '(LAMBDA$ (I) (* I I))
                                    (BINARY-* I I)))
               (1 2 3 4 5))
  <2 (COLLECT$ (1 4 9 16 25))
<1 (ACL2_*1*_ACL2::COLLECT$ (1 4 9 16 25))
(1 4 9 16 25)
ACL2 !>
\end{verbatim}

%%% [From Matt] I replaced the following two lines by the line just
%%% below them, because it may seem to suggest that warrants are not
%%% required, yet badges are indeed required.
% Top-level evaluation pays no attention to warrants (or warrant
% hypotheses), because they are all true at the top level.  In short,
Top-level evaluation treats all warrants as being true at the top level.  In short,
the reason is that at the top level, attachments may be used (see :DOC
\hrefdoc{ACL2\_\_\_\_DEFATTACH}{defattach}), and all warrants have
true attachments.  Thus for example, if we introduce the function
symbol \texttt{square} as defined in Section~\ref{sec:warrants} and we
replace the body of the \loopd expression above by \texttt{(square
  i)}, then we get the same evaluation result.

\subsection{Guards and fast Common Lisp evaluation for function calls}\label{sec:guards}

% [J agrees] With my change regarding warrants and the Special
% Conjectures, are we done worrying about saying something about
% warrants in guards?

Recall that \hrefdocu{ACL2\_\_\_\_GUARD}{guard} verification permits
ACL2 to execute with Common Lisp definitions.  When this is done,
\loopd expressions are computed by evaluating corresponding Common
lisp \loopc expressions.  For example, recall this definition from
Section~\ref{sec:warrants}.

\begin{verbatim}
(defun f2 (lower upper)
  (declare (xargs :guard (and (integerp lower) (integerp upper))))
  (loop$ for i of-type integer from lower to upper
         collect (square i)))
\end{verbatim}

\noindent If we evaluate \texttt{(f2 1 5)}, we get the same result as
in the preceding example: \texttt{(1 4 9 16 25)}.  However, if we
first trace \collectd, as before, then this time we see no calls of
\collectd.  The reason is clear from the single-step Common Lisp
macroexpansion of the \loopd expression displayed below, which shows
that when attachments are allowed (i.e., \texttt{*aokp*} is true), as
is the case when evaluating directly in the top-level loop, then the
corresponding \loopc expression is evaluated.  (We'll discuss the
other case when we discuss evaluation during proofs.)

\begin{Verbatim}[commandchars=\\\{\}]
(COND (*AOKP* (LOOP FOR I OF-TYPE INTEGER FROM LOWER TO UPPER
                    COLLECT (SQUARE I)))
      (T ...))
\end{Verbatim}

Guard verification is necessary before a \loopd call will be evaluated
using \loopc in Common Lisp.  Indeed, if we instead define \texttt{f2}
by adding the \hrefdocutt{ACL2\_\_\_\_XARGS}{xargs} declaration
\texttt{:verify-guards nil}, then evaluation of \texttt{(f2 1 5)} will
show \collectd calls in a trace as displayed in the preceding
subsection.  The natural question is then: What are the guard proof
obligations generated for a \loopd expression?

Let us begin by considering our \texttt{f2} example.  In that case,
ACL2 reports the following ``non-trivial part of the guard
conjecture''.

\begin{verbatim}
(IMPLIES (AND (INTEGERP LOWER)
              (INTEGERP UPPER)
              (APPLY$-WARRANT-SQUARE)
              (MEMBER-EQUAL NEWV (FROM-TO-BY LOWER UPPER 1)))
         (INTEGERP NEWV))
\end{verbatim}

\noindent This formula states that if \texttt{newv} is a member of the
indicated \texttt{from-to-by} expression, then \texttt{newv} is an
integer.  It is generated by the expression ``\texttt{i of-type
  integer}'' in the \loopd expression, which is necessary for guard
verification of the loop body, since the guard of \texttt{(square n)}
is \texttt{(integerp n)}.

The formula displayed above illustrates the first of the following
three classes of guard proof obligations arising from \loopd
expressions.  The second one below is a rather obvious requirement.
The third arises because of Common Lisp quirks: all tails may be
checked in the \texttt{ON} case, including \nil, and the type-checks
for indices from \texttt{i} to \texttt{j} by \texttt{k} may include
one step past the point where iteration stops.

\begin{itemize}

\item {\em Special Conjecture (a)}: Every member of the target list
  satisfies the guard of the function object.

\item {\em Special Conjecture (b)}: On every member of the target
  list, the function object produces a result of the right type: a
  number if the loop operator is \kSUM and a true list if the operator
  is \kAPPEND.

\item {\em Special Conjectures (c)}: For \texttt{(\loopd $\ldots$ ON
  $\ldots$)}, \nil satisfies the type expression (if any).  For a
  \texttt{(FROM-TO-BY i j k)} target list: \texttt{i}, \texttt{j},
  \texttt{k}, and \texttt{(+ i (* k (floor (- j i) k)) k)} each
  satisfy the type expression (if any).

\end{itemize}

Warrant hypotheses can be relevant to guard verification, although the
user is generally spared from thinking about that.  Consider the
following variant of \texttt{f2} above, where \kSUM replaces
\kCOLLECT.

\begin{verbatim}
(defun sum-squares-2 (lower upper)
  (declare (xargs :guard (and (integerp lower) (integerp upper))))
  (loop$ for i of-type integer from lower to upper
         sum (square i)))
\end{verbatim}

\noindent The hypothesis \texttt{(warrant square)} does not appear in
the \texttt{:guard}, but it is added automatically in the generated
guard conjecture.  This addition is necessary; the proof fails without
it.  This addition is also justified: the Special Conjectures are
generated only to avoid guard violations when evaluating the Common
Lisp \loopc expression that corresponds to the \loopd expression, and
that only happens when attachments are permitted (see the use of
\texttt{*aokp*} in the Common Lisp definition of \loopd above).  But
when attachments are permitted then all warrant hypotheses are true,
which justifies them as hypotheses.

We conclude this subsection by noting that instead of using the
\texttt{of-type} construct, one can supply a \texttt{:guard} as
indicated in Section~\ref{sec:syntax}.  The \loopd expression in the
body of \texttt{f2} could instead be written as follows, giving an
equivalent definition.
\begin{verbatim}
(loop$ for i from lower to upper
       sum :guard (integerp i) (square i))
\end{verbatim}
The \texttt{of-type} construct may provide for more efficient Common
Lisp code, since the \texttt{:guard} construct is ignored by the
compiler.  On the other hand, the \texttt{:guard} construct is more
flexible, since it can reference more than one variable.

\subsection{Evaluation during proofs}\label{sec:eval-proofs}

An example in Section~\ref{sec:reasoning} shows that evaluation of
\loopd expressions can take place during proofs.  That example did not
require any warrant hypotheses, but in general they might be required.

As noted above, when attachments are allowed then all warrant
hypotheses hold, so ACL2 evaluates \texttt{(\applyd\ '$F$ $\ldots$)}
by simply calling $F$ on the supplied arguments.  Unfortunately,
attachments are not allowed during proofs.

However, a major design goal for ACL2 is fast evaluation during
proofs.  That includes calls of \applyd, calls of \loopd, and calls of
functions that lead to calls of \applyd or \loopd.  Thus, a challenge
is for ACL2 to support evaluation of \texttt{(\applyd\ '$F$ $\ldots$)}
by simply calling $F$, as discussed above for the case that
attachments are allowed.

ACL2 meets this challenge, as clearly illustrated by its ability to
prove the following theorem using evaluation of a \loopd expression
(where \texttt{sum-squares-2} is defined as in the preceding
subsection).

\begin{verbatim}
(thm (implies (warrant square)
              (equal (sum-squares-2 1 4) 30)))
\end{verbatim}

\noindent Let us inspect the summary.

\begin{verbatim}
Rules: ((:DEFINITION SUM-SQUARES-2)
        (:EXECUTABLE-COUNTERPART EQUAL)
        (:EXECUTABLE-COUNTERPART FROM-TO-BY)
        (:EXECUTABLE-COUNTERPART SUM$)
        (:REWRITE APPLY$-SQUARE))
\end{verbatim}

\noindent The use of the rule \texttt{(:rewrite apply\$-square)} is
expected, as discussed in Section~\ref{sec:reasoning}.  The use of the
definition rule for \texttt{sum-squares-2} (as opposed to evaluation
of that function's call) is perhaps surprising: it takes place early,
in the {\em preprocess} step of the
\hrefdocu{ACL2\_\_\_\_HINTS-AND-THE-WATERFALL}{waterfall}, where
evaluation of \loopd expressions that require warrants is not
supported.  The simplifier then evaluates the body of
\texttt{sum-squares-2} with respect to the binding alist
\texttt{((upper . '4) (lower . '1))}.  The hints shown below avoid the
definition rule
but the proof still succeeds by running the executable counterpart of
\texttt{sum-squares-2}, which actually checks that the necessary warrant is
available as a hypothesis.

\begin{verbatim}
(thm (implies (warrant square)
              (equal (sum-squares-2 1 4) 30))
     :hints (("Goal" :in-theory (disable sum-squares-2))))
\end{verbatim}

Deleting the hypothesis causes the proof to fail after the missing warrant
is forced by proof-time evaluation.

For more examples involving evaluation of \loopd expressions during proofs,
see the \hrefdoc{ACL2\_\_\_\_COMMUNITY-BOOKS}{community books}
\texttt{system/tests/apply-in-proofs.lisp} and
\texttt{system/tests/loop-tests.lisp}.

We conclude our discussion of evaluation during proofs by remarking on
the implementation and its impact on performance.  In the preceding
subsection we discussed the macroexpansion of \loopd expressions in
Common Lisp when attachments are allowed.  We now consider the other
case, which invokes the usual translation to a call of a \loopd scion.
The expansion is as follows for the \loopd expression in the
definition of \texttt{f2}.

\begin{Verbatim}[commandchars=\\\{\}]
(COND (*AOKP* {\em{; Attachments are allowed.}}
       (LOOP FOR I OF-TYPE INTEGER FROM LOWER TO UPPER
             COLLECT (SQUARE I)))
      (T (COLLECT$ '(LAMBDA (I)
                            (DECLARE (TYPE INTEGER I)
                                     (XARGS :GUARD (INTEGERP I) :SPLIT-TYPES T)
                                     (IGNORABLE I))
                            (RETURN-LAST 'PROGN
                                         '(LAMBDA$ (I)
                                                   (DECLARE (TYPE INTEGER I))
                                                   (SQUARE I))
                                         (SQUARE I)))
                   (FROM-TO-BY LOWER UPPER '1))))
\end{Verbatim}

\noindent The \loopd scion call is generated by finding a term
associated with the \loopd expression in the
\hrefdocu{ACL2\_\_\_\_WORLD}{world}.

\begin{verbatim}
ACL2 !>(cdr (assoc-equal '(LOOP$ FOR I OF-TYPE INTEGER
                                 FROM LOWER TO UPPER COLLECT (SQUARE I))
                          (global-val 'LOOP$-ALIST (w state))))
(LOOP$-ALIST-ENTRY
     (COLLECT$ '(LAMBDA (I)
                        (DECLARE (TYPE INTEGER I)
                                 (XARGS :GUARD (INTEGERP I)
                                        :SPLIT-TYPES T)
                                 (IGNORABLE I))
                        (RETURN-LAST 'PROGN
                                     '(LAMBDA$ (I)
                                               (DECLARE (TYPE INTEGER I))
                                               (SQUARE I))
                                     (SQUARE I)))
               (FROM-TO-BY LOWER UPPER '1)))
ACL2 !>
\end{verbatim}

\noindent ACL2 updates the \texttt{loop\$-alist} structure at the
conclusion of a \texttt{defun} event (here, for \texttt{f2}), by
collecting all {\em tagged loops} \texttt{(RETURN-LAST 'PROGN '(LOOP\$
  ...) $e$)} and associating each such \texttt{(LOOP\$ ...)} with its
corresponding translation, $e$.  Well, almost: the problem is that $e$
might not be executable, in particular because
\hrefdocutt{ACL2\_\_\_\_MV-LET}{mv-let} calls are eliminated by
translation using \hrefdocutt{ACL2\_\_\_\_MV-NTH}{mv-nth} calls.
So the implementation (with function
\texttt{convert-tagged-loop\$s-to-pairs}) converts the \loopd
translation to executable code (with function
\texttt{logic-code-to-runnable-code}), notably by inserting calls of
\hrefdocutt{ACL2\_\_\_\_MV-LIST}{mv-list}.

A more complete answer also takes into account early loading of
compiled files and removes ACL2-specific guard information.  Here is
the definition of \loopd (from the ACL2 sources) in Common Lisp.

\begin{verbatim}
(defmacro loop$ (&whole loop$-form &rest args)
  (let ((term (or (loop$-alist-term loop$-form
                                    *hcomp-loop$-alist*)
                  (loop$-alist-term loop$-form
                                    (global-val 'loop$-alist
                                                (w *the-live-state*))))))
    `(cond (*aokp* (loop ,@(remove-loop$-guards args)))
           (t ,(or term
                   '(error "Unable to translate loop$ (defun given directly ~
                            to raw Lisp?)."))))))
\end{verbatim}

The following two issues are raised by this method of evaluating
\loopd expressions during proofs.

\begin{itemize}

\item Section~\ref{sec:warrants} noted the necessity of warrant
  hypotheses because attachments are disallowed during proofs.
  Specifically, as we have seen, calls of \loopd scions may ultimately
  lead to calls of \applyd on user-defined functions, which is where
  warrant hypotheses are required.  The implementation permits
  evaluation of such \applyd calls during rewriting but
  \hrefdocu{ACL2\_\_\_\_FORCE}{force}s the necessary warrant
  hypotheses that are not known in the current rewriting context (the
  so-called ``\hrefdocu{ACL2\_\_\_\_TYPE-ALIST}{type-alist}'').

\item Evaluation of \loopd expressions can be considerably slower
  during proofs than it is at the top level, because of the use of
  \loopd scions and \applyd rather than evaluation of a Common Lisp
  \loopc expression.  We considered allowing this faster evaluation
  during proofs, but that would require checking (or forcing) the
  warrant hypothesis of {\em every} user-defined function in the loop
  body (and also the \kWHEN and \kUNTIL clauses, if present), even for
  functions that are rarely called when evaluating the loop body, such
  as in the error branch of an \texttt{IF} expression.

\end{itemize}

\subsection{Performance}

%% !! Run these numbers again using Version_8.2 when it comes out.

The results reported in the comments below were produced using an ACL2
version (git hash 5eb79e7697) built on CCL on April 4, 2018, running
on a 3.5 GHz 4-core Intel(R) Xeon(R) with Hyper-Threading.  Times in
seconds are realtime; also shown are bytes allocated.  The key
take-away is the comparison between (c) and (d): Evaluation in the
ACL2 loop, including the \texttt{integer-listp} call in the guard, is
very close in time to the corresponding evaluation directly in Common
Lisp.

\begin{verbatim}
(include-book "projects/apply/top" :dir :system)
(defun$ double (n)
  (declare (xargs :guard (integerp n)))
  (+ n n))
(defun sum-doubles (lst)
  (declare (xargs :guard (and (integer-listp lst)
                              (warrant double))
                  :verify-guards nil))
  (loop$ for x of-type integer in lst sum (double x)))
(make-event `(defconst *m* ',(loop$ for i from 1 to 10000000 collect i)))
; (a) ACL2 top-level loop$ call [0.98 seconds, 160,038,272 bytes]:
(time$ (loop$ for i of-type integer in *m* sum (double i)))
; (b) ACL2 top-level non-guard-verified function call [0.89 seconds, 160M bytes]
(time$ (sum-doubles *m*))
(verify-guards sum-doubles)
; (c) ACL2 top-level guard-verified function call [0.14 seconds, 16 bytes]
(time$ (sum-doubles *m*))
(value :q)
; (d) Common Lisp guard and function call [0.13 seconds, 0 bytes]:
(time$ (and (integer-listp *m*) (sum-doubles *m*)))
; (e) Common Lisp function call [0.09 seconds, 0 bytes]:
(time$ (sum-doubles *m*))
; (f) Common Lisp loop call [0.08 seconds, 0 bytes]:
(time$ (loop for i of-type integer in *m* sum (double i)))
\end{verbatim}

\section{Limitations, Future Work, and Conclusion}\label{sec:conclusion}

We know of several limitations that may be addressed in future work.
For more information search for ``Possible Future Work on Loop\$'' in
ACL2 source file \texttt{translate.lisp}.

\begin{itemize}

%%% Added by Moore 2 Oct, 2019:
\item Reliance on \applyd and scions means that quoted \texttt{lambda}
  objects may appear in formulas that, when entered by the user, had no such
  objects in them.  This can cause trouble because two obviously equivalent
  quoted \texttt{lambda} objects may be distinct objects.  This arises
  especially when a theorem statement may involve a \loopd iteration variable
  with a different name than used in some function definition.  E.g.,
  \texttt{'(lambda (x) (sq x))} is a different object than \texttt{'(lambda
    (y) (sq y))} even though they are obviously functionally equivalent.
  \texttt{Declare} forms in such objects can also render them unnecessarily
  distinct.  While the functional equivalence of \texttt{lambda} objects is,
  of course, undecidable, trivial cases like these ought to be caught by the
  prover but are not.  Furthermore, because they are so obvious the user may
  overlook the differences!  Until these issues are smoothed out in the
  prover, checkpoints involving \texttt{lambda} objects must be read with
  unusual care!

\item \Applyd only works on \hrefdocu{ACL2\_\_\_\_LOGIC}{logic} mode
  functions so it is an error if a
  \hrefdocu{ACL2\_\_\_\_PROGRAM}{program} mode function appears in a
  \loopd body or a \kWHEN or \kUNTIL clause.

\item \Applyd does not work on \hrefdocu{ACL2\_\_\_\_STATE}{state}- or
  \hrefdocu{ACL2\_\_\_\_STOBJ}{stobj}-using functions, hence neither
  does \loopd.  This may be very difficult to change.

\item At the time this paper was submitted, recursion within a \loopd body
  was not supported and so it is not described here.  But since then we have
  added support for it; see :DOC
  \hrefdocu{ACL2\_\_\_\_LOOP\_42-RECURSION}{loop\$-recursion}.

\item As noted at the end of Subsection~\ref{sec:eval-proofs},
  evaluation of a \loopd expression uses \loopd scions instead of the
  more efficient Common Lisp \loopc when either during a proof or
  directly in the top-level loop (rather than in a function body).
  (Of course, such evaluation can be expected to be much faster than
  term rewriting.)  For the top-level case, perhaps ACL2 should report
  an error and insist on the use of the utility
  \hrefdocutt{ACL2\_\_\_\_TOP-LEVEL}{top-level}, as illustrated by the
  following example.

\begin{verbatim}
ACL2 !>(time$ (loop$ for i from 1 to 10000000 sum i))
; (EV-REC *RETURN-LAST-ARG3* ...) took
; 1.33 seconds realtime, 1.34 seconds runtime
; (320,039,824 bytes allocated).
50000005000000
ACL2 !>(time$ (top-level (loop$ for i from 1 to 10000000 sum i)))
50000005000000
; (EV-REC *RETURN-LAST-ARG3* ...) took
; 0.05 seconds realtime, 0.05 seconds runtime
; (235,648 bytes allocated).
ACL2 !>
\end{verbatim}

\item Common Lisp \loopc supports more general forms than \loopd, some
  of which are feasible to support in \loopd.  Here are a couple of
  examples.

\begin{verbatim}
? (loop for x in '(2 20 5 50 3 30) by #'cddr maximize x)
5
? (loop for i from 11/2 downto 1 by 2 collect i)
(11/2 7/2 3/2)
? 
\end{verbatim}

\end{itemize}

\noindent Despite these limitations, we have seen that \loopd provides
efficient execution and can make reasoning more succinct.  We expect
to evolve its implementation as users tell us what most needs
improvement.

\section*{Acknowledgments}

This material is based upon work supported in part by DARPA under
Contract No. FA8650-17-1-7704.  We are also grateful for support of
this work by ForrestHunt, Inc., and for helpful reviewer feedback.
% (From Matt) Strictly speaking I think the comma after ``Inc.'' above
% isn't quite right grammatically.  But I think it enhances
% readability.

\bibliographystyle{eptcs}
\bibliography{paper.bib}
\end{document}